\documentclass[twocolumn]{aastex61}
\pdfoutput=1 %for arXiv submission
\usepackage{amsmath,amstext}
\usepackage[T1]{fontenc}
\usepackage[figure,figure*]{hypcap}
\usepackage{color}

\def\PAK{PA$_{kin}$}
\def\ATLAS3D{ATLAS$^{3D}$}

\shorttitle{Kinematic alignment of cluster ETGs}
\shortauthors{Kim et al.}

\begin{document}

\title{Discovery of the kinematic alignment of early-type galaxies in the Virgo cluster}

\author{Suk Kim}
\author{Hyunjin Jeong}
\affiliation{Korea Astronomy $\&$ Space Science institute, Daejeon 34055, Republic of Korea; star4citizen@kasi.re.kr}
\author{Jaehyun Lee}
\affiliation{Korea Institute for Advanced Study, 85, Hoegi-ro, Dongdaemun-gu, Seoul 02455, Republic of Korea}
\author{Youngdae Lee}
\author{Seok-Joo Joo}
\author{Hak-Sub Kim}
\affiliation{Korea Astronomy $\&$ Space Science institute, Daejeon 34055, Republic of Korea; star4citizen@kasi.re.kr}
\author{Soo-Chang Rey}
\affiliation{Department of Astronomy and Space Science, Chungnam National University, Daejeon 34134, Republic of Korea}
%\author{Author B}
%\affiliation{Affiliation 1}
%\affiliation{Affiliation 3}
%\author{Author C}
%\affiliation{Affiliation 3}
%\author{Author D}
%\affiliation{Affiliation 4}

\begin{abstract}
Using the kinematic position angles (\PAK), an accurate indicator for the spin axis of a galaxy, obtained from the \ATLAS3D\ integral-field-unit (IFU) spectroscopic data, we discovered that 57 Virgo early-type galaxies tend to prefer the specific \PAK\ values of 20\arcdeg\ and 100\arcdeg, suggesting that they are kinematically aligned with each other. These kinematic alignment angles are further associated with the directions of the two distinct axes of the Virgo cluster extending east-west and north-south, strongly suggesting that the two distinct axes are the filamentary structures within the cluster as a trace of infall patterns of galaxies. Given that the spin axis of a massive early-type galaxy does not change easily even in clusters from the hydrodynamic simulations, Virgo early-type galaxies are likely to fall into the cluster along the filamentary structures while maintaining their angular momentum. This implies that many early-type galaxies in clusters are formed in filaments via major mergers before subsequently falling into the cluster. Investigating the kinematic alignment in other clusters will allow us to understand the formation of galaxy clusters and early-type galaxies.
\end{abstract}

\keywords{Galaxy: kinematics and dynamics -- galaxies: clusters: individual (Virgo) -- galaxies: formation}

\section{Introduction} 
  The alignment between galaxies and the cosmic web is an important factor for comprehensively understanding structure formation in the Universe. Of particular interest is the spin axes of early-type galaxies, because one expects them to exhibit specific patterns in filaments related to the cosmic web. Numerical simulations, for example, have shown that redder and more massive galaxies in large-scale structures tend to have spin axes perpendicular to the filaments direction \citep{Codis2012,Dubois2014}. These simulated results agree with observations \citep{Tempel2013a,Tempel2013b}. 

  In cluster environments, on the other hand, previous studies have investigated the alignment of galaxies in various aspects. Among them, two of the most popular alignments are (1) the brightest cluster galaxy (BCG) alignment, and (2) satellite alignment. The former is the alignment between the major axis of the BCG and the direction of the elongation of the cluster. This BCG alignment can be explained by filamentary accretion processes and the tidal field effect by the dark matter halo of the cluster \citep{Dubinski1998,Faltenbacher2008}. \cite{Huang2016} found a detection of the BCG alignment in 8237 clusters in the redshift range of 0.1 $<$ z $<$ 0.35, and \cite{West2017} confirmed this phenomenon in more distant clusters (0.19 $<$ z $<$ 1.8). The latter is the alignment of the major axes of the cluster member galaxies toward the cluster center. This satellite alignment is considered to originate from the tidal torque induced by the strong gravitational field of the galaxy cluster, which can change the shape of the member galaxies. It has been observationally confirmed by several studies \citep{Pereira2005,Agustsson2006,Faltenbacher2007}, but some authors have claimed that they could not detect the satellite alignment in their studies \citep[see, e.g.,][]{Bernstein2002,Siverd2009,Schneider2013,sifon2015}.
 
  Another approach to understanding the alignment of galaxies in cluster environments is to compare the spin axes of member galaxies with the cluster structure.
In $N$-body simulations, \cite{Tormen1997} showed that infalling galaxies are distributed with the shape of the host dark matter halo, and \cite{Knebe2004} further claimed that infalling galaxies preserve the direction of orbits when they first fall into the cluster. The shape of the host dark matter halo thus would be related to the infall pattern of galaxies. One of the prominent features of the Virgo cluster, the dynamically young cluster \citep{Arnaboldi2004,Aguerri2005}, is the presence of two distinct axes \citep{Binggeli87,schindler1999}. If these two distinct axes within the Virgo cluster are a part of filaments around the cluster as a signature of infall patterns \citep{Kim2016},
%\textcolor{red}{1: the alignment between the two axes and Virgo early-type galaxies could be found.}
the spin axes of Virgo early-type galaxies would be perpendicular to the directions of the two distinct axes.

In this Letter, we focus on the spin axes of Virgo early-type galaxies based on the kinematic position angles (\PAK) from the \ATLAS3D\ integral-field-unit (IFU) spectroscopic survey, which can provide information on the exact angular momentum vector. This study will give insight into the formation of the cluster and early-type galaxies. 
%\ylee{3: the alignment between spin axes of galaxies and the direction of the filament can be found.}
%\ylee{2: To catch insight into the formation of the cluster and early-type galaxies, in this study, we focus on the spin axes of Virgo early-type galaxies based on the kinematic position angles (\PAK) from the \ATLAS3D\ integral-field-unit (IFU) spectroscopic survey, which can provide information on the exact angular momentum vector.}

\section{DATA}
  To investigate the alignments of cluster early-type galaxies, we used the value of \PAK\ from the \ATLAS3D\ IFU survey \citep{Krajnovi2011}, a volume-limited sample of 260 early-type galaxies in the local universe \citep{Cappellari2011}. The \PAK\ is defined as the angle between the north and the receding part of the velocity map in a counterclockwise direction, which is, in principle, perpendicular to the spin axis. 
In this paper, however, the values of \PAK\ from \ATLAS3D\ are rearranged to be in the range of 0\arcdeg\ to 180\arcdeg\ by not considering whether the receding or preceding parts of the velocity map. That is, if the \PAK\ is 210\arcdeg\ in \ATLAS3D, it is assigned to 30\arcdeg\ in this Letter (see inset of Figure~\ref{fig:fig1}).

In previous studies on the alignments between the spin axes of galaxies and large-scale structures, the spin axis of a galaxy was estimated from the photometric position angle. If a galaxy has a very small ellipticity or prolate rotation, however, its photometric position angle will be not sufficiently accurate to estimate the spin axis. Therefore, the \PAK\ from the IFU observation is an important factor in determining more accurately the direction of the spin axis, especially for early-type galaxies with small ellipticities. 
Note that 57 out of the 260 \ATLAS3D\ galaxies corresponding to certain members in the EVCC \citep[Extended Virgo cluster Catalog;][]{Kim2014} are classified as "$Virgo$", while the others (203) are "$non$-$Virgo$". Our {\it Virgo} sample contains all massive early-type galaxies (M$_{k}<$$-$21.6) in the Virgo cluster. 

%\textcolor{red}{To measure the alignment angle regardless of the spin direction, we redefine \PAK\ as increasing counterclockwise from the north (see inset of Figure~\ref{fig:fig1})}.
%\textcolor{green}{To measure the alignment angle do not regarding opposite spin direction. Therefore, we redefine \PAK\ as increasing counterclockwise from the north-pole until 90\arcdeg(The increasing clockwise from the north-pole is defined as a negative value until $-$90\arcdeg; see inset of Figure~\ref{fig:fig1}).}
%\ylee{When securing the \PAK\ from \ATLAS3D\ , we redefine \PAK\ as increasing counterclockwise from the north (see inset of Figure~\ref{fig:fig1}) considering the alignment angle regardless of the spin direction.} 

\section{Results}
\subsection{Distribution of early-type galaxies in the Virgo cluster}

\begin{figure}
\centering
  \includegraphics[width=3.4in]{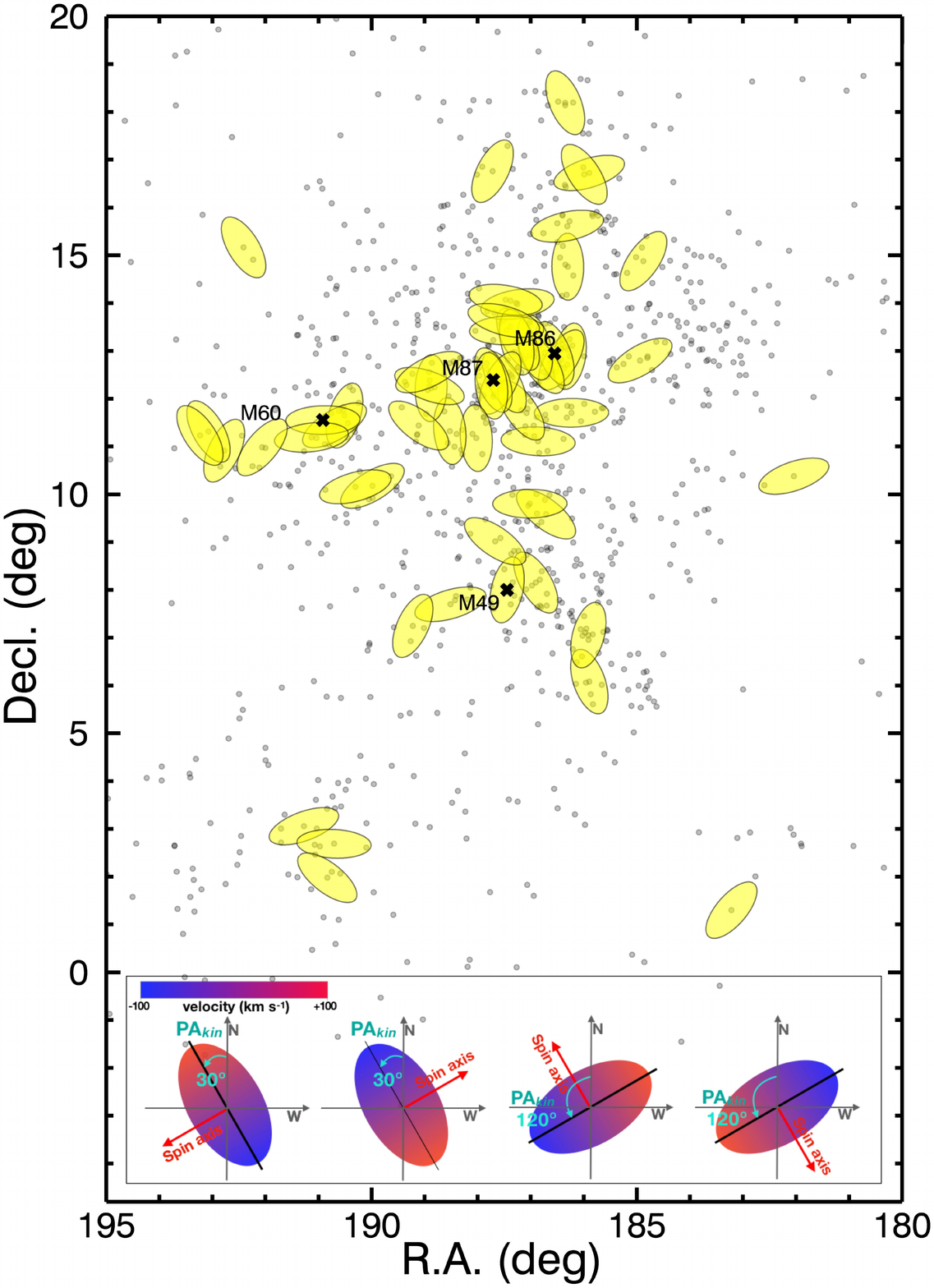}
  \caption{%Projected spatial distribution of galaxies in the Virgo cluster. The gray dots are certain member galaxies in the EVCC. The yellow ellipses indicate 57 early-type galaxies in the \ATLAS3D\ sample among certain members of the EVCC. For the ellipses, the directions of the major axes point to the \PAK\ of the galaxies, and their sizes are given arbitrarily. The inset shows a schematic of \PAK.
  Projected spatial distribution of galaxies in the Virgo cluster. The yellow ellipses indicate 57 $Virgo$ early-type galaxies from the \ATLAS3D\ sample, while the gray dots are certain member galaxies of the Virgo cluster from the Extended Virgo Cluster Catalog (EVCC). For the ellipse, the major axis comes from \PAK, and the size is given arbitrarily. The inset shows a schematic diagram of \PAK.}
  \label{fig:fig1}
\end{figure}

  The Virgo cluster has not only very complex structure composed of two major subclusters (centered on the giant early-type galaxies, M87 and M49) and many subgroups, but also the anisotropic distribution of massive early-type galaxies represented by two distinct axes \citep{Binggeli87,schindler1999,West2000,Mei2007}. One axis stretches east-west across M60-M87-M86, where massive elliptical galaxies in the central region of the Virgo cluster are aligned in three dimension. 
This east-west axis \citep[hereafter the principal axis according to][]{West2000} is suggested to be connected with Leo II A and B filaments around the Virgo cluster \citep{Kim2016}.
The other axis extends north-south, which traces a chain of galaxies and groups including the M49 group and massive early-type galaxies such as NGC 4365, NGC 4261, and NGC 4342. This north-south axis is also visible in the residual X-ray image obtained by the subtraction of the spherically symmetric $\beta$-model image \citep[see figure 2 of][]{Bohringer1994}. The largest M49 sub-group is known to be falling into the cluster through this axis and will probably flow into the cluster center \citep{Kraft2011}. 
The anisotropic galaxy distribution in a cluster, such as the two distinct axes in the Virgo cluster, is naturally predicted by cold dark matter hierarchical clustering models and this distribution may be related to large-scale structures \citep{Knebe2004}.  
These two distinct axes, therefore, might be the filamentary structure within the cluster made by infalling galaxies.

In Figure~\ref{fig:fig1}, we present the projected spatial distribution of 57 $Virgo$ \ATLAS3D\ early-type galaxies as yellow ellipses together with EVCC certain member galaxies as gray dots. The major axis of each ellipse comes from the \PAK\ of the galaxy. Our 57 massive early-type galaxies are located along the two distinct axes, confirming the previous studies. 

\subsection{Alignment Signal}

\begin{figure*}[!htbp]
\centering \includegraphics[width=6in]{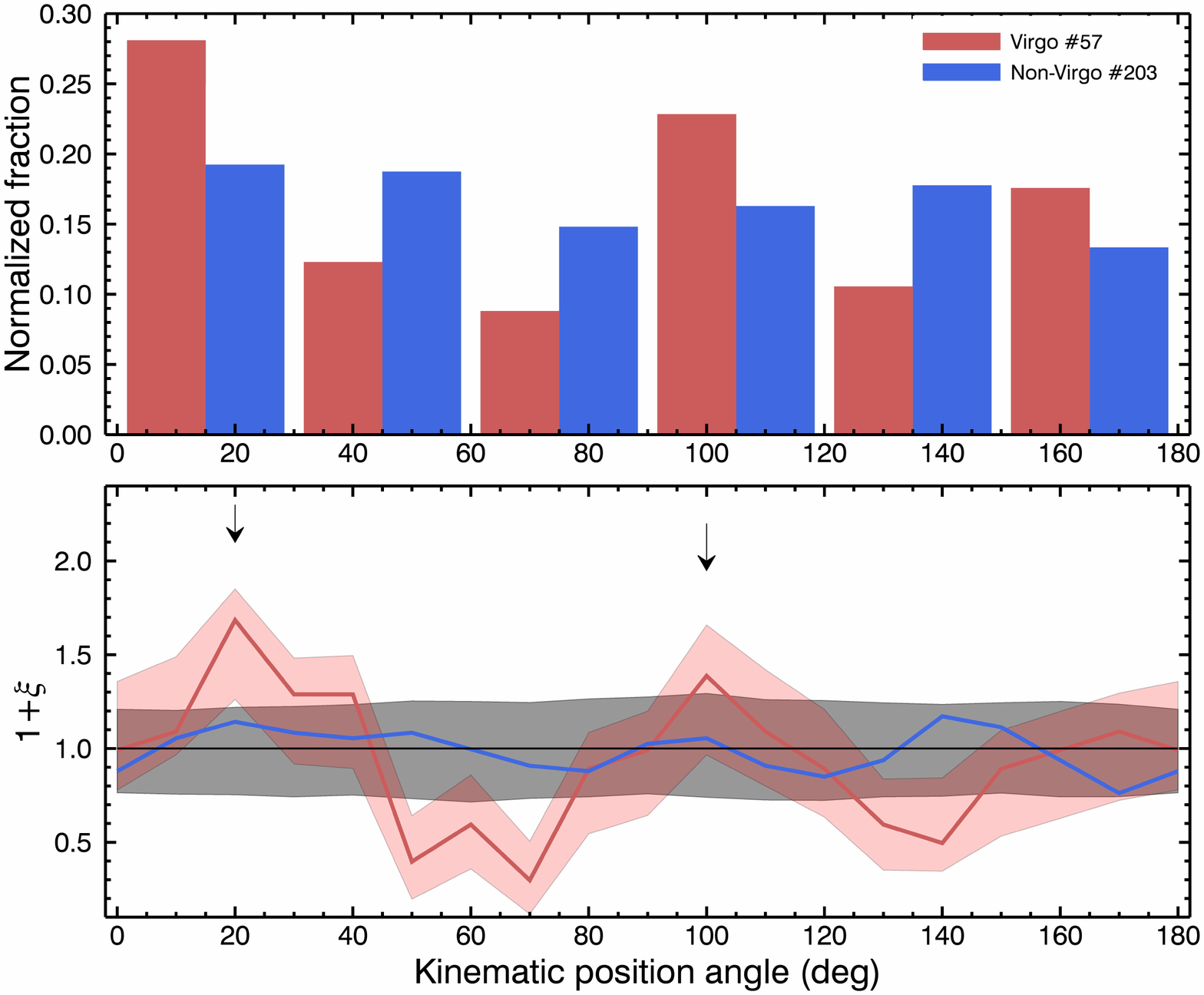}
  \caption{%Top: Kinematic position angle (\PAK) distribution of galaxies. The bin width is 30\arcdeg. The red and blue bars represent $Virgo$ and $non-Virgo$, respectively. Bottom: Probability distribution function as a running histogram with a bin width of 30\arcdeg\ and an incremental step of 10\arcdeg. The red and blue solid lines are $Virgo$ and $non-Virgo$, respectively. The red and black shaded regions indicate the 1$\sigma$ confidence interval of $Virgo$ and uniform samples, respectively. The 1$\sigma$ confidence interval was obtained by 1000 bootstrap resampling test. The black arrows mark the alignment angles of $Virgo$.
  Top: kinematic position angle (\PAK) distribution of galaxies with a bin width of 30\arcdeg. The red and blue histograms represent the $Virgo$ and $non-Virgo$ samples, respectively. Bottom: Probability distribution function as a running histogram with a bin width of 30\arcdeg\ and an incremental step of 10\arcdeg. The red, blue, and black solid lines are $Virgo$, $non-Virgo$ and uniform samples, respectively, while the red and black shaded regions indicate the 1$\sigma$ confidence interval of $Virgo$ and uniform samples, respectively. The 1$\sigma$ confidence interval was obtained by 1000 bootstrap resampling test. The black arrows mark the alignment angles of $Virgo$.}
  \label{fig:fig2}
\end{figure*}

  We present, for the first time, the \PAK\ distribution of 57 $Virgo$ \ATLAS3D\ early-type galaxies in the top panel of Figure~\ref{fig:fig2}. The $Virgo$ early-type galaxies (red histogram) prefer specific values for \PAK\, roughly 20\arcdeg\ and 100\arcdeg, while the distribution of \PAK\ for the $non$-$Virgo$ galaxies (blue histogram) is relatively uniform. This implies that some Virgo early-type galaxies could be kinematically aligned with each other in terms of \PAK.

  To confirm this alignment, we calculated the probability distribution function (PDF; 1+$\xi$) as a running histogram with a bin width of 30\arcdeg\ and an incremental step of 10\arcdeg, where $\xi$ is the excess probability of \PAK\ in the bins. The bottom panel of Figure~$\ref{fig:fig2}$ presents the PDFs of \PAK\ for $Virgo$ (PDF$_{Virgo}$, red line) and $non$-$Virgo$ (PDF$_{non-Virgo}$, blue line). The PDF$_{Virgo}$ exhibits two peaks at about 20\arcdeg\ and 100\arcdeg, marked as black arrows, while the PDF$_{non-Virgo}$ exhibits an almost uniform distribution of 1+$\xi$ = 1. We performed the Kuiper and Watson tests\footnote{These circular statistical tests are carried out using the $R$ statistical package (\url{http://www.r-project.org/})} to quantify the statistical significance of the differences between the PDF$_{virgo}$ and null hypothesis distribution that represents the random distribution. Both tests give a probability of $<$ 0.01 rejecting the null hypothesis. %On the other hand, the Kuiper and Watson tests for the PDF$_{non_Virgo}$ yield probabilities of $>$ 0.85 and 0.99, respectively.
  
%  We performed the Kolmogorov-Smirnov (K-S) test to quantify the statistical significance of the difference between the PDF$_{virgo}$ and the uniform distribution (PDF$_{uniform}$; black line in bottom panel of Figure~\ref{fig:fig2}). The test gives a  probability of $<$0.01 rejecting the null-hypothesis that the two samples are drawn from the same parent population. On the other hand, the K-S test for the PDF$_{non-Virgo}$ compared to the PDF$_{uniform}$ yields a probability of 0.41. 

  As an additional step, to statistically verify that our finding is not caused by the small sample size, we performed 1000 bootstrap resamplings for both $Virgo$'s \PAK\ and the uniform distribution with the same number ($\#$57) considering the error propagation. The red and black shaded regions correspond to a 1$\sigma$ confidence interval from the bootstrapping for $Virgo$ ($\sigma_{e{\_}Virgo}$) and the uniform sample ($\sigma_{e{\_}uniform}$), respectively. Note that the amplitude fluctuation of the PDF$_{Virgo}$ is larger than about 5.3$\sigma_{e{\_}Virgo}$, while that of PDF$_{non-Virgo}$ is buried in $\sigma_{e{\_}uniform}$. We thus confirm that the two peaks at about 20\arcdeg\ and 100\arcdeg\ (hereafter the kinematic alignment angles) are significant even considering the errors and found that over half of our $Virgo$ early-type galaxies (37/57) are in the range of  0\arcdeg\,$<$\,\PAK\,$<$\,40\arcdeg\ or 80\,\arcdeg\,$<$\,\PAK\,$<$\,120\arcdeg. This implies that many Virgo early-type galaxies are kinematically aligned with each other for these two different kinematic alignment angles. 

%\textcolor{blue}{As an additional step, we performed 1000 bootstrap resampling test in order to statistically verify that our finding is not caused by the small sample size: employing $Virgo$'s \PAK\ and the uniform distribution as parent samples, respectively, we randomly selected a subsample with the same number (i.e., 57).} 

\begin{figure}
 \centering \includegraphics[width=3.3in]{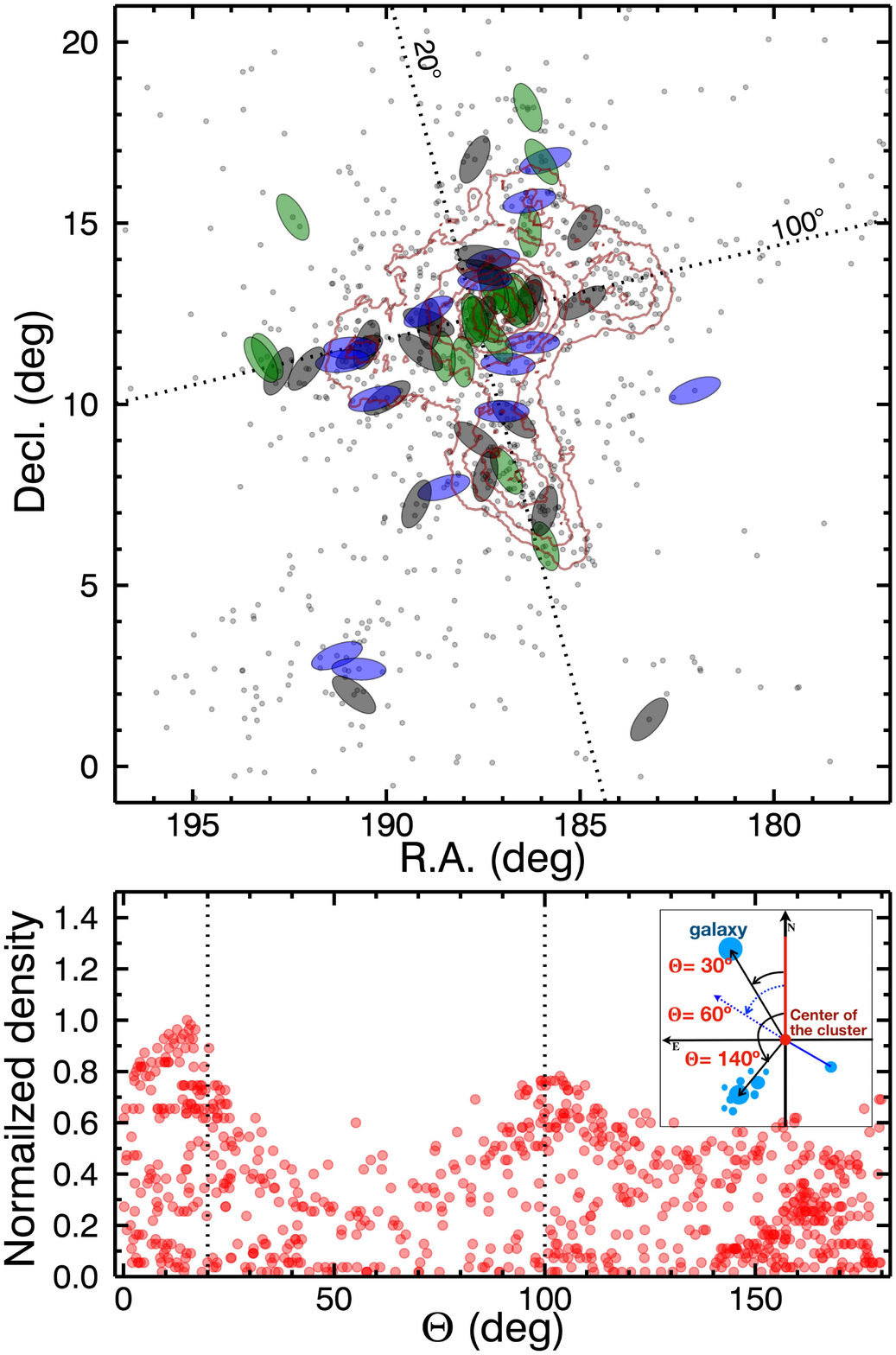}
  \caption{Top: Same as Figure~\ref{fig:fig1} but with different ellipse colors. The green and blue ellipses indicate $Virgo$ early-type galaxies in two ranges of \PAK\ of 20\arcdeg\ (0\arcdeg\,$<$\,\PAK\,$<$\,40\arcdeg) and 100\arcdeg (80\arcdeg\,$<$\,\PAK\,$<$\,120\arcdeg), while the gray ellipses are the remaining galaxies. The contours denote the two-dimensional local density of galaxies. Bottom: Local density as a function of separation angle ($\Theta$) from the North Pole in the Virgocentric reference frame. The inset shows a schematic diagram of separation angle. The filled circle represents the local density of a galaxy calculated from the number of galaxies within 1\arcdeg\ radius circle. The red filled circle represents represents a galaxy located outside 2\arcdeg\ from the cluster center.}
  \label{fig:fig3}
\end{figure}

\subsection{Alignment angle and structure of the Virgo cluster}

  The top panel of Figure~\ref{fig:fig3} presents the local density distribution of galaxies in the Virgo cluster (red contours). The local density for each galaxy was calculated by counting the number of galaxies inside the 1\arcdeg\ radius circle (including that galaxy) based on the EVCC certain members. The contours extend east-west and north-south from the center of the cluster (position of M87) following the two distinct axes.
Interestingly, the kinematic alignment angles (i.e., \PAK\, 20\arcdeg\ and 100\arcdeg, dotted lines) closely correspond to the directions of the two distinct axes of the Virgo cluster. 
That is, galaxies with 0\arcdeg\,$<$\,\PAK\,$<$\,40\arcdeg\ (green ellipses) and 80\,\arcdeg\,$<$\,\PAK\,$<$\,120\arcdeg\ (blue ellipses) are generally located along the north-south (20\arcdeg) axes and the principal (100\arcdeg), respectively.

  The bottom panel of Figure~\ref{fig:fig3} presents the local density distribution of galaxies as a function of their separation angles, $\Theta$,  from the North Pole in the Virgocentric reference frame. To find the structure penetrating the cluster center, in this analysis, we set the range of the $\Theta$ to be from 0\arcdeg to 180\arcdeg\ (see inset in bottom panel of Figure~\ref{fig:fig3}). Furthermore, We excluded galaxies within 2\arcdeg\ from the cluster center where the cluster structure could be strongly affected by dynamical relaxation \citep{Lapi2011}.
The distribution of the remaining galaxies, marked as red filled circles, clearly shows that galaxies with high local density are concentrated on the two separation angles around 20\arcdeg\ and 100\arcdeg\ (vertical dotted lines) exactly tracing the two distinct axes, which are identical to the kinematic alignment angles.

The coincidence between the kinematic alignment angles and the directions of the two distinct axes suggests that the origin of the $Virgo$ early-type galaxies is strongly connected to the two distinct axes, which are expected to be the filamentary structures within the cluster.

%The coincidence between the kinematic alignment angles and the directions of the two distinct axes
%\color{red}confirms that the two axes are the filamentary structures through which the $Virgo$ early-type galaxies fell into the cluster.\color{black}

\section{Conclusions}
  To understand the formation of clusters and early-type galaxies, we investigated the kinematic alignment of 57 $Virgo$ early-type galaxies based on \PAK\ from the \ATLAS3D\ IFU survey and found that they prefer specific \PAK\ values, roughly 20\arcdeg\ and 100\arcdeg. These two kinematic alignment angles are further linked to the directions of the filamentary structures within the Virgo cluster. What, then, is the origin of these kinematic alignments in the Virgo cluster?

  In filaments, several observational studies \citep{Codis2012,Tempel2013a,Tempel2015,Tempel2013b,Hirv2017} indicate that the major axes (perpendicular to spin axes) of early-type galaxies are expected to be aligned with the direction of the host filaments because major mergers that form early-type galaxies occur along the filament, making their major axes parallel to the directions of the filaments. Furthermore, in filaments, central galaxies in groups tend to be early-type galaxies compared with galaxies outside of the filaments \citep[e.g.,][]{Kuutma2017,Poudel2017}. This suggests that the filament environment could efficiently transform the galaxy morphology from late- to early-type. 
  
  According to the hydrodynamic simulation by J. Lee et al. (2018, in preparation), furthermore, the spin axis of an early-type galaxy does not change easily even after they fall into the cluster. If early-type galaxies in the cluster come from the filaments maintaining the angular momentum, their \PAK\ would be aligned with each other, and we might find the infalling pattern such as filamentary structures.
  
  %without the interaction between galaxies such as mergers. 
  
%If early-type galaxies in filaments fall into the cluster, we might find \color{black}the infalling pattern such as filamentary structures and these early-type galaxies could be aligned their \PAK\ with the direction of the filamentary structure.

%\textcolor{orange}{If early-type galaxies in filaments fall into the cluster, we might find the infalling pattern such as filamentary structures, since their \PAK\ would be aligned with each other.}

%\textcolor{blue}{If early-type galaxies in filaments fall into the cluster, their \PAK\ would be aligned with the direction of the filamentary structure, and we might find the infalling pattern such as filamentary structures.}

%we might find \color{black}the infalling pattern such as filamentary structures and these early-type galaxies could be aligned their \PAK\ with the direction of the filamentary structure.

 % We thus conclude that many cluster early-type galaxies are already formed via major mergers along the filament before falling into the cluster. 
%\textcolor{red}{Our results suggested that many early-type galaxies in the Virgo cluster are already formed via major mergers along the filament, after then falling into the cluster maintaining their direction of angular momentum.}
Our results suggest that many Virgo early-type galaxies are likely to be already formed via major mergers along the filaments before falling into the cluster.
%maintaining their direction of angular momentum.}
By investigating other clusters of various dynamical stages,
%we need to prove whether our findings are common in all cluster early-type galaxies. 
%\textcolor{red}{we will know our findings are common or not in general.}
we will know whether our findings are common or not.
This would clarify the build-up mechanism of galaxy clusters characterized by the accretion of galaxies along filaments.

\acknowledgments
%The authors acknowledges support from the National Junior Research Fellowship of National Research Foundation(NRF) of Korea (No. 2011-0012618) and the Basic Science Research Programs through the NRF funded by the Ministry of Education, Science, and Technology (NRF-2013R1A6A3A04064993, 2015R1A2A2A01006828 and 2018R1A2B2006445). Support for this work was also provided by the NRF to the Center for Galaxy Evolution Research (2017R1A5A1070354). 
The authors acknowledges support from the National Junior Research Fellowship of National Research Foundation (NRF) of Korea (No. 2011-0012618) and the Basic Science Research Programs through the NRF funded by the Ministry of Education, Science, and Technology (NRF-2013R1A6A3A04064993, 2015R1A2A2A01006828 and 2018R1A2B2006445). Support for this work was also provided by the NRF to the Center for Galaxy Evolution Research (2017R1A5A1070354).

%This research was supported by the Basic Science Research Program through the National Research Foundation of Korea (NRF) funded by the Ministry of Education, Science, and Technology (2015R1A2A2A01006828). 

%\facility{facility ID}
%\facilities{facility ID, facility ID, facility ID} 
%\software{Numpy}

%\bibliographystyle{yahapj}
%\bibliography{references}

\end{document}